\documentclass{aa}
\usepackage{graphicx}
\usepackage{txfonts}  
\usepackage{psfrag}
\usepackage{natbib}
\usepackage{verbatim}
\usepackage{amssymb}
\usepackage{latexsym}
\newcommand{\kref}[1]{{\S} \ref{#1}}

\begin{document}

\title{\object{GRB 021004}: Tomography of a gamma-ray burst progenitor and 
    its host galaxy
    \thanks {Based on observations taken with the ESO\' \rm s 8.2m Very Large Telescope in Chile.}  
}  
  
\author{A. J.       Castro-Tirado      \inst{1} 
   \and P.          M\o ller           \inst{2}
   \and G.          Garc\'{\i}a-Segura \inst{3}
   \and J.          Gorosabel          \inst{1}
   \and E.          P\'erez            \inst{1}
   \and A.          de Ugarte Postigo  \inst{4}
   \and E.          Solano             \inst{5}
   \and D.~         Barrado-Navascu\'es\inst{5}
   \and S.          Klose              \inst{6}
   \and D. A.       Kann 	       \inst{6}
   \and J. M.       Castro Cer\'on     \inst{7}
   \and C.          Kouveliotou        \inst{8}
   \and J. P. U.    Fynbo              \inst{9}
   \and J.          Hjorth             \inst{9}
   \and H.          Pedersen           \inst{9}
   \and E.          Pian               \inst{2,10,11}
   \and E.          Rol                \inst{12,13}
   \and E.          Palazzi            \inst{14}
   \and N.          Masetti            \inst{14}
   \and N. R.       Tanvir             \inst{12}
   \and P. M.       Vreeswijk          \inst{9}
   \and M. I.       Andersen           \inst{9}
   \and A. S.       Fruchter           \inst{15}
   \and J.          Greiner            \inst{16}
   \and R. A. M. J. Wijers             \inst{13}
   \and E. P. J.    van den Heuvel     \inst{13}
      } 
 
\offprints{A.J. Castro-Tirado, \email{ajct@iaa.es} }  
 
\institute{Instituto de Astrof\'\i sica de Andaluc\'\i a (IAA-CSIC), Glorieta de la Astronom\'{\i}a s/n, E-18.008 Granada, Spain. 
       \and European Southern Observatory, Karl-Schwarzschild-Straße 2, 85748, Garching bei M¨unchen, Germany. 
	 \and Instituto de Astronom\'{\i}a, Universidad Nacional Aut\'onoma de M\'exico, Apdo. Postal 877, Ensenada 22800, Baja California, M\'exico. 
	\and INAF, Osservatorio Astronomico di Brera, via E. Bianchi 46, 23807 Merate (LC), Italy. 
       \and Laboratorio de Astrof\'{\i}sica Estelar y Exoplanetas, Dpto. Astrof\'{\i}sica, Centro de Astrobiolog\'{\i}a (CSIC/INTA), P.O. Box 78, 28691 Villanueva de la Ca\~nada (Madrid), Spain. 
       \and Th\"uringer Landessternwarte Tautenburg, Sternwarte 5, D-07778 Tautenburg, Germany.
       \and European Space Agency (ESA), European Space Astronomy Centre (ESAC), P.O. Box - Apdo. de correos 78, 28691 Villanueva de la Ca\~nada, Madrid, Spain.
        \and NASA Marshall Space Flight Center, NSSTC, 320 Sparkman Drive, Huntsville, Alabama 35805, USA.
       \and Dark Cosmology Centre, Niels Bohr Institute, University of Copenhagen, Juliane Maries Vej 30, DK-2100 K{\o}benhavn {\O}, Denmark.
       \and INAF - Osservatorio Astronomico di Trieste, via Tiepolo, 11, 34131 Trieste, Italy.
       \and Scuola Normale Superiore di Pisa, Piazza dei Cavalieri 7, I-56126 Pisa, Italy.
       \and Department of Physics \& Astronomy, University of Leicester, University Road, Leicester, LE1 7RH, UK.
       \and Astronomical Institute ``Anton Pannekoek'', University of Amsterdam, PO number 94249, 1090 GE, Amsterdam, NL.
       \and INAF - IASF di Bologna, via Gobetti 101, I-40129 Bologna, Italy.
       \and Space Telescope Science Institute, 3700 San Mart\'{\i}n Dr, Baltimore, MD~~21218-2463, USA.
       \and Max-Planck-Institut f\"ur extraterrestrische Physik, 85748 Garching, Germany.
           } 
 
\date{Received 24 December 2009; accepted 26 March 2010}

\abstract
  {}
  {Analyse the distribution of matter around the progenitor star of gamma-ray burst GRB 021004 as well as the properties of its host galaxy with high-resolution echelle as well as near-infrared spectroscopy.
   }
  {Observations were taken by the 8.2m Very Large Telescope with the Ultraviolet and Visual Echelle spectrograph (UVES) and the Infrared Spectrometer And Array Camera (ISAAC) 
  between 10 and 14 hours after the onset of the event.}
  {We report the first detection of emission lines from a GRB host galaxy in the near-infrared, detecting H$\alpha$ and the [O {\small  III}] doublet. These allow an 
   independent measurement of the systemic redshift ($z=2.3304\pm0.0005$) which is not contaminated by absorption as the Ly$\alpha$ line is, and the deduction of properties 
   of the host galaxy.  
   From the visual echelle spectroscopy, we find several absorption line groups spanning a range 
   of about 3,000 km s$^{-1}$ in velocity relative to the redshift of the host 
   galaxy. The absorption profiles are very complex with both 
   velocity-broadened components extending over several 100 km s$^{-1}$ and 
   narrow lines with velocity widths of only $\sim$ 20 km s$^{-1}$.
   By analogy with QSO absorption line studies, the relative velocities,
   widths, and degrees of ionization of the lines (``line-locking'',
   ``ionization--velocity correlation'') show that the progenitor had both 
   an extremely strong radiation field and several distinct mass loss phases 
   (winds).} 
  {These results are consistent with GRB progenitors being massive stars, 
   such as Luminous Blue Variables (LBVs) or Wolf--Rayet stars, providing a 
   detailed picture of the spatial and velocity structure of the GRB 
   progenitor star at the time of explosion.
   The host galaxy is a prolific star-forming galaxy with a SFR of
   $\sim$40 M$_{\odot}$yr$^{-1}$.}


\maketitle  
  
\section{Introduction}  

The afterglows of long-duration Gamma-Ray Bursts (GRBs), which are linked with the
explosions of massive stars \citep[see][for a recent review]{woo06}, are the
most luminous optical sources in the universe for short periods of time
\citep{kan07,blo09}. Low-resolution optical spectroscopy was initially only
usable to determine the redshift and thus place them at cosmological distances
\citep{met97}. Deeper insight came with the first medium-resolution spectrum,
obtained with Keck ESI of the afterglow of GRB 000926 \citep{cas03}. The first
true high-resolution echelle spectra were obtained for GRB 020813 \citep{fio05},
but they were of low signal-to-noise ratio. The first echelle spectra with good
S/N were finally obtained for GRB 021004, the focus of this work \citep[see also
][]{fio05}.

Such spectroscopy allows deep insight into the environments of GRBs \citep{pro06}.
Some highlights include the possible detection of a Galactic superwind in the 
host galaxy of GRB 030329 \citep{tho07}, and variable absorption lines which
result from direct UV pumping by the luminous GRB afterglow \citep{des06, vre07,
del09}. These detections have recently been possible for more afterglows due to
the rapid localization capabilities of the \emph{Swift} satellite \citep{geh04}
in combination with the rapid-response mode (RRM) which is now available for the
Ultraviolet-Visual Echelle Spectrograph (UVES) at the Very Large Telescope (VLT)
\citep[e.g.,][]{vre07,del09}. Recently, the covered wavelength region has been
expanded all the way from the ultraviolet into the $K$ band near-infrared (nIR)
by the second-generation instrument X-Shooter at VLT \citep[][see \citealt{deu09} for a first
result]{dod06}. 

GRB 021004   was detected at   12:06:14 universal time  (UT)  on 2002,
October 4 with the gamma-ray  instrument FREGATE, the wide-field X-ray
Monitor (WXM) and  the soft X-ray  camera (SXC) aboard the High-Energy
Transient Explorer (\emph{HETE-2}) \citep{shi02}. \object{GRB~021004},
was  a moderately  bright, long-duration  ($T_{90}=100$ s) event  with
fluences  of   $6.4\times10^{-7}$   erg   cm$^{-2}$ (7-30     keV) and
$2.3\times10^{-6}$ erg cm$^{-2}$ (30-400 keV) \citep{bar02}.

The GRB was rapidly localized in flight and  the position was reported
in  less than a minute,  allowing   rapid  ground-based follow-up  which
revealed the  presence of the fading optical   afterglow of GRB 021004
\citep{fox03}.   This    prompted  follow-up  observations   at   many
observatories, which led to  an extensive long-term coverage  at 
X-ray \cite{sak02a, sak02b}, radio
\citep{fra02, ber02, poo02a, poo02b}\footnote{See http://www.aoc.nrao.edu/$\sim$dfrail/grb021004.dat for the complete VLA data set.},  
millimeter \citep{deu05}, near-IR  \citep{deu05,fyn05}
and      optical    wavelengths,          both            ground-based
\citep{fox03,ber03,uem03,hol03,pan03,mir03,kaw04,deu05}  as  well   as
space-based \citep{fyn05}.

The isotropic  energy release during  the prompt emission of  this GRB
was    modest,    with   log    $E_{iso}=52.65^{+0.12}_{-0.17}$
\citep{kan10}, and the dust extinction  for this somehow reddish afterglow
(R-K $\sim$ 3) was also low as is typical for
many well-observed  afterglows \citep{kan06}.  Despite  the low energy
promptly released,  this is  among the  most luminous  afterglows  ever detected
\citep{kan06}, even in comparison  with a much larger \emph{Swift}-era
sample  \citep{kan10}. The multiwavelength  temporal evolution  of the
GRB 021004  afterglow can be  explained by multiple  energy injections
\citep{bjo04, deu05},  nevertheless  other  scenarios  can not  be  discarded
\citep{laz02}.   GRB  021004 remains  one  of  the most  well-observed
afterglows ever.

Early-time  low  and   medium  resolution  spectroscopic  observations
allowed a  redshift  of  $z=2.33$ to  be  determined on  the  basis of
Ly-$\alpha$ absorption   and  emission  lines \citep{cho02}.   Several
absorption systems with outflows velocities  of  few 1000 km  s$^{-1}$
were   also  reported  \citep{sal02,sav02} and  studied  in  detail by
\citet{mol02b}, \cite{wan03}, \citet{mat03}, \citet{sch03}, \citet{mir03}, \citet{sta05} and \cite{laz06}.
\cite{jak05} also report the detection of the host galaxy Ly$\alpha$ line
in narrow-band imaging.

\citet{fio05} have constrained the ionization parameters
of the various absorption  components detected above Ly$\alpha$ (at an
observer  restframe of 4,050  \rm \AA) in GRB 021004,  interpreting them  as density
fluctuations, like \citet{laz02} did. Within the context of larger samples, the UVES spectra
of this GRB have also been studied by \cite{che07} (the general lack of
wind signatures in high-resolution spectra of GRB afterglows), \cite{pro08}
(N {\small V} absorption lines toward GRB afterglows), \cite{fox08} (high-ionisation line systems toward GRB afterglows) and \cite{tej07, tej09, ver09}
(study of Mg {\small II} foreground absorption systems).

The original expectation, as detailed high quality spectroscopy of
GRB afterglows became possible, was that we would rapidly learn
much about the GRB progenitors via the study of the complex
absorption line systems they were expected to exhibit due to
ejection events leading up to the final collapse. This expectation
has far from proven true.

Despite the large number of long GRB afterglow spectra obtained since,
GRB 021004 still stands out as the one with the most complex
set of intrinsic (0-3,000 km/s ejection velocity) absorption systems.
At the time it also held the place as the lowest detected HI column
density, and it still ranks between the lowest seven found.
The complexity originally suggested that it indeed represented
a display of ejecta \citep{mol02a, mir03, fio05} but this interpretation 
was later disputed by \citet{che07}. A final interpretation of the 
complex systems has not yet been agreed on, and this object still stands 
our as the best candidate of a seemingly rare case where the signatures of
events prior to the collapse are displayed. The rarity alone
would warrant a more detailed discussion of the spectral
features. In addition we are adding UVES data below 4050 ang
for a more detailed discussion of the absorption systems
(para 3.1) and near-IR VLT/ISAAC spectroscopy allowing a more
accurate determination of host properties and redshift (para 3.2). 
We discuss our results in the light of progenitor models for GRBs in 
\kref{Discussion}, and summarize the work in \kref{Conclusions}.

\begin{figure*}[t!]  
\begin{center} 
  \resizebox{15.2cm}{!}{\includegraphics[clip]{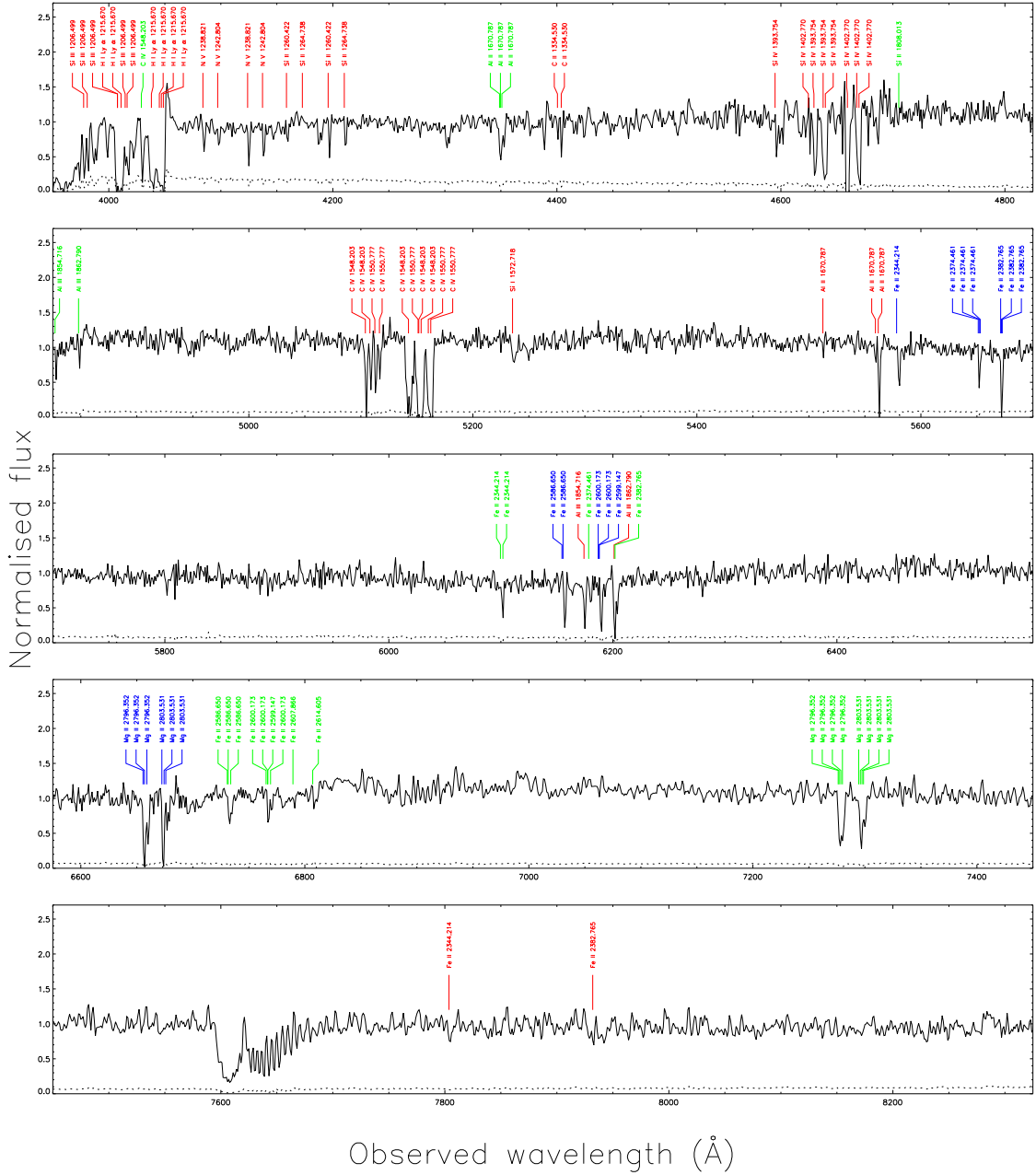}}  
 
      \caption{Overall   view of    GRB  021004    optical   afterglow
      spectrum. These VLT/UVES data were obtained $\sim$0.6 days after
      the GRB  and show the Ly$\alpha$  emission line arising from the
      host galaxy redshift plus some  of the most prominent absorption
      lines systems at (or very  close  to) the host galaxy  redshift:
      low ionization  lines (like Fe  {\small II}, Si {\small II}) and
      high ionization lines  (like Si {\small  IV},  C {\small IV},  N
      {\small    V}), all  labelled   in red   colour. Also shown  for
      completeness are  the   absorption  systems for  the  foreground
      systems at $z  = 1.6020$  (labelled in green)  and $z  = 1.3820$
      (labelled in  blue). For clarity,  the  original data have  been
      smoothed. The telluric   lines in the  A-band (7600--7630 \AA{})
      are noticeable. The dotted line is the error spectrum.}  
      \label{uves spectrum}
\label{SpectrumFull}
\end{center}  
\end{figure*} 

For a Hubble constant of H$_{0}$ = 72 km s$^{-1}$ Mpc$^{-1}$, a matter
density $\Omega_{m}$ = 0.3, and a cosmological constant 
$\Omega_{\Lambda}$ = 0.7, the luminosity distance to the host is 
d$_{L}$ = 18.22 Gpc, and the look-back time is 10.42 Gyr. 
All errors are given at a $1\sigma$ level of confidence for a parameter 
of interest unless stated otherwise.

\section{Observations and data reduction}  
  \label{observaciones}  
    
 \subsection{Optical observations} Observations were conducted    with
 8.2-m Very Large Telescope Units   1 (VLT/UT1, Antu) and 2  (VLT/UT2,
 Kueyen) at the European  Southern Observatory (ESO) in  Cerro Paranal
 (Chile).  We  obtained  optical  spectroscopy of the    GRB afterglow
 starting 0.6~days  after    the burst using  the   Ultraviolet-Visual
 Echelle Spectrograph (UVES) on  UT2. Contemporaneous nIR observations
 were taken  using the Infrared Spectrometer  And Array Camera (ISAAC)
 on UT1 (see Table \ref{ObsTable} for a log of the observations).

The optical data reduction was performed using the UVES context
running under MIDAS\footnote{MIDAS (Munich Imaging Data Analysis System) is
developed and maintained by the European Southern Observatory (ESO). www.eso.org/projects/esomidas/midas-distrib.html}.
The UVES context is structured into reduction recipes allowing for
data reduction  in a semi-automatic manner.
A master bias frame was  used for the bias subtraction. The order tracing was
performed taking  advantage of a  first guess calibration solution based on the
UVES physical model. In all  cases, except for the red spectrum centered at
8,600 {\AA} which is affected by fringing effects, the flat field correction
was applied after the order extraction. A Th-Ar lamp was used for wavelength
calibration.  The spectra were extracted using an optimal
extraction method, which provides the object signal and the variance. After the
extraction the spectra were  resampled to constant wavelength bins and the
orders merged into a single spectrum. Finally, both the standard heliocentric
and vacuum wavelength corrections were applied.

\subsection{Near-infrared observations}
  For the nIR observations  we used the short  wavelength mode in  low
  resolution, which yields a spectral coverage of 1.42-1.83 $\mu$m and
  1.84-2.56 $\mu$m in the H and K bands, respectively. The pixel scale
  of  the   ESO-Hawaii detector is  0\farcs146/pixel.  For
  each band, 50  individual exposures of 60  seconds each  were taken,
  shifting the target along the slit every 5 exposures in H (3 in  K) 
  in order to  be able to remove the sky  background, a  
  standard technique in  infrared astronomy. We also collected
  dark and flat field images, which were used in the reduction process
  performed  with IRAF\footnote{IRAF    is the  Image   Reduction  and
  Analysis  Facility, a   general  purpose  software system  for   the
  reduction and  analysis  of astronomical  data. IRAF  is written and
  supported  by  the IRAF  programming  group  at the National Optical
  Astronomy Observatories (NOAO)   in Tucson, Arizona,   USA.}.  After
  combining the  groups of exposures    and extracting the  individual
  spectra, we  calibrated  them using the  air-glow  OH  emission lines
  present   in  the 2  dimensional  images.   This  allows an accurate
  wavelength calibration with dispersion of  4 and 8  {\AA}/pix for H
  and   K  respectively.     Finally,  the   individual  spectra  were
  median-combined into  a   single one.   In  the case  of  the H band
  spectrum we  used  a G2 spectral   type star to  remove the telluric
  absorption  bands. The spectra were  flux calibrated by means of the
  JHK-band  observations   that   we   obtained   bracketing   the spectral
  observations.

\begin{table}[t!]
\begin{minipage}[t]{\columnwidth}
\begin{center}
\caption{Journal of the VLT GRB~021004 optical/nIR spectroscopic observations. All observations were obtained on October 5 (UT).}
\begin{tabular}{clcccc}
\noalign{\smallskip} \hline\hline \noalign{\smallskip}
UT Time  & Instr. & Exp.     & Spectral  &  S/N   & Resolution\\
(start)  &            & time (s) & Range (\rm \AA) & ratio\footnote{per resolution element.} & $\lambda/\Delta(\lambda)$\\
\hline
03:02  &  UVES  &  3600    &  3290-4520    &  1   & $\sim$ 56000 \\
       &        &           &  4620-5595    & 2.2  & $\sim$ 51500 \\
       &        &           &  5675-6645    & 2.7  & $\sim$ 51500 \\
03:02  &  ISAAC &  3600    &  14200-18300  & 5.2  & $\sim$ 500    \\
04:10  &  UVES  &  3600    &  3050-3870    &  1   & $\sim$ 51300 \\
       &        &           &  4780-5755    & 3.7  & $\sim$ 50500 \\
       &        &           &  5835-6805    & 5.4  & $\sim$ 50500 \\
05:15  &  UVES  &  3600    &  3750-4650    & 2.4  & $\sim$ 51500 \\
       &        &           &  6705-8520    & 4.1  & $\sim$ 52300 \\
       &        &           &  8665-10400   & 2.2  & $\sim$ 48700 \\
05:27  &  ISAAC &  3000    &  18400-25800  & 2.3  & $\sim$ 450    \\
\noalign{\smallskip} \hline \hline\end{tabular}
\label{ObsTable}
\end{center}
\end{minipage}
\end{table}

\begin{table}[t!]
\begin{minipage}[t]{\columnwidth}
\begin{center}
\caption{Individual high-velocity metal-lines systems in GRB 021004.
}
\scriptsize{
\renewcommand{\footnoterule}{}
\begin{tabular}{lccccc}
\noalign{\smallskip} \hline\hline \noalign{\smallskip}
 Line ID                  & C2      &    C1   &    D     &     E2   &    E1  \\
\noalign{\smallskip} \hline \noalign{\smallskip}
Si III $\lambda$1,206     &    y \footnote{y = yes; $-$ = no; ? = doubtful, low significance due to low S/N ratio in the S IV region.}  &    y    &    y     &     y    &    y   \\
H I Ly$\alpha$ $\lambda$1,216 &   y &    y    &    y     &     y    &    y   \\
N V $\lambda$1,238-1,242  &    y    &  $-$    &   $-$    &   $-$    &    y   \\
Si II $\lambda$1,260-1,264&  $-$    &    y    &    ?     &   $-$    &    y   \\
C II   $\lambda$1,334     &    y    &    y    &   $-$    &   $-$    &   $-$  \\
Si IV $\lambda$1,394-1,403&    ?    &    ?    &    ?     &     ?    &    y   \\
C IV $\lambda$1,548-1,551 &    y    &    y    &    y     &     y    &    y   \\
Al II $\lambda$1,671      &   $-$   &    y    &   $-$    &     y    &    y   \\
Al III $\lambda$1,855-1,863&  $-$   &  $-$    &   $-$    &   $-$    &    y   \\
Fe II $\lambda$2,344-2,382 &  $-$   &  $-$    &   $-$    &   $-$    &    y   \\
\hspace{1.35cm} $z$        &2.29671 & 2.29935 &  2.3216  &  2.3275  & 2.32891\\
\hspace{1.35cm} v (km/s)\footnote{relative to the systemic velocity of the host galaxy at z = 2.3304, which we set as zero.} & $-$3,050\footnote{Embedded as C1 in the wide C complex (2,730-3,250 km s$^{-1}$).} & $-$2,810\footnote{Two subcomponents at $-$2,837 and $-$2,803 km s$^{-1}$ (low ionization) and $-$2,860 and $-$2,810 km s$^{-1}$ (high ionization).} & $-$795\footnote{ Wide complex spanning 585-1,005 km s$^{-1}$.} & $-$262\footnote{Embedded as E1 in the wide E complex (40-390 km s$^{-1}$).} & $-$134\footnote{Two subcomponents at $-$150 and $-$114 km s$^{-1}$.}  \\
\noalign{
\smallskip} \hline \hline\end{tabular}
} \normalsize \rm
\label{SystemsTable}
\end{center}
\end{minipage}
\end{table}

\begin{table}[ht!]
\begin{minipage}[t]{\columnwidth}
\begin{center}
\caption{Line Identifications in the GRB 021004 OA spectra.}
\tiny{
\renewcommand{\footnoterule}{}
\begin{tabular}{lccc}
\noalign{\smallskip} \hline\hline \noalign{\smallskip}
$\lambda_{vacuum}$(observed) (\rm \AA)     & Line ID     & 
EW (\rm \AA)\footnote{Gaussian profile fitting (for non saturated 
lines only). Star symbols refer to complex systems.}       & z \\
\hline
   4008       &     H I Ly$\alpha$ 1215.670  &             &  2.2967\\
   4015       &     Si III 1206.499          &             &  2.3275\\
   4016       &     Si III 1206.499          &             &  2.3289\\
   4028       &     C IV   1548.203          &             &  1.6024\\
   4038       &     H I Ly$\alpha$ 1215.670  &             &  2.3216\\
   4041       &     H I Ly$\alpha$ 1215.670  &             &  2.3243\\
   4045       &     H I Ly$\alpha$ 1215.670  &             &  2.3275\\
   4047       &     H I Ly$\alpha$ 1215.670  &             &  2.3289\\
   4049\footnote{Emission line. Partially absorbed, Ly$\alpha$ flux =
(1.5 $\pm$ 0.2) $\times$ 10$^{-16}$ erg cm$^{-2}$ s$^{-1}$.} &  H I Ly$\alpha$ 1215.670  &             &  2.3304\\
   4084.0     &     N V    1238.821    &  0.46 $\pm$ 0.16  &  2.2967\\
   4097.2     &     N V    1242.804    &  0.35 $\pm$ 0.15  &  2.2967\\
   4123.9     &     N V    1238.821    &  0.90 $\pm$ 0.14  &  2.3289\\
   4137.2     &     N V    1242.804    &  0.66 $\pm$ 0.12  &  2.3289\\
   4158.5     &     Si II  1260.422    &  0.25 $\pm$ 0.10  &  2.2993\\
   4195.8     &     Si II  1260.422    &  0.71 $\pm$ 0.19  &  2.3289\\
   4210.2     &     Si II  1264.738    &  0.51 $\pm$ 0.16  &  2.3289\\
   4347.4     &     Al II  1670.787    &  0.49 $\pm$ 0.17  &  1.6020\\
   4399.6     &     C II   1334.530    &  0.38 $\pm$ 0.15  &  2.2968\\
   4403.0     &     C II   1334.530    &  0.79 $\pm$ 0.14  &  2.2994\\
   4639.7     &     Si IV  1393.754    &  1.39 $\pm$ 0.83  &  2.3289\\
   4669.7     &     Si IV  1402.770    &  1.29 $\pm$ 0.46  &  2.3289\\
   5104.0     &     C IV   1548.203    &  1.77 $\pm$ 0.12  &  2.2967\\
   5107.8     &     C IV   1548.203    &  0.16 $\pm$ 0.03  &  2.2992\\
   5108.5     &     C IV   1548.203    &  0.84 $\pm$ 0.10  &  2.2998\\
   5112.5     &     C IV   1550.777    &  1.15 $\pm$ 0.09  &  2.2967\\
   5116.2     &     C IV   1550.777    &  0.06 $\pm$ 0.04  &  2.2992\\
   5117.0     &     C IV   1550.777    &  0.67 $\pm$ 0.08  &  2.2998\\
   5138.3     &     C IV   1548.203    &                   &  2.3189\\
   5140.0     &     C IV   1548.203    &                   &  2.3200\\
   5143.1     &     C IV   1548.203    &                   &  2.3220\\
   5146.7     &     C IV   1550.777    &                   &  2.3188\\
   5146.7     &     C IV   1548.203    &                   &  2.3243\\
   5148.5     &     C IV   1550.777    &                   &  2.3200\\
   5148.5     &     C IV   1548.203    &                   &  2.3255\\
   5151.7     &     C IV   1550.777    &                   &  2.3220\\
   5152.9     &     C IV   1548.203    &                   & 2.3275+2.3289  \\
   5155.2     &     C IV   1550.777    &                   &  2.3243\\
   5157.6     &     C IV   1550.777    &                   &  2.3255\\
   5161.5     &     C IV   1550.777    &                   & 2.3275+2.3289  \\
   5235.4     &     Si I   1572.718    & 0.80 $\pm$ 0.10   &  2.3289\\
   5559.5     &     Al II  1670.787    & 0.23 $\pm$ 0.07   &  2.3275\\
   5561.9     &     Al II  1670.787    & 1.73 $\pm$ 0.12   &  2.3289\\
   5578.0     &     Fe II  2344.214    & 1.01 $\pm$ 0.15   &  1.3796      \\
   5651.2     &     Fe II  2374.461    & 0.52 $\pm$ 0.06   &  1.3800      \\
   5652.4     &     Fe II  2374.461    & 0.16 $\pm$ 0.05   &  1.3805      \\
   5652.9     &     Fe II  2374.461    & 0.13 $\pm$ 0.02   &  1.3807      \\
   5671.2     &     Fe II  2382.765    & 0.86 $\pm$ 0.07   &  1.3801      \\
   5671.9     &     Fe II  2382.765    & 0.41 $\pm$ 0.06   &  1.3804      \\
   5672.6     &     Fe II  2382.765    & 0.37 $\pm$ 0.06   &  1.3807      \\
   6099.6     &     Fe II  2344.214    & 0.59 $\pm$ 0.04   &  1.6020      \\
   6101.7     &     Fe II  2344.214    & 0.28 $\pm$ 0.04   &  1.6029      \\
   6154.5     &     Fe II  2586.650    & 0.67 $\pm$ 0.08   &  1.3794      \\
   6155.0     &     Fe II  2586.650    & 0.50 $\pm$ 0.09   &  1.3798      \\
              &     Fe II  2365.552    & 0.50 $\pm$ 0.09   &  1.6020      \\
   6174.2     &     Al III 1854.716    & 1.15 $\pm$ 0.09   &  2.3289      \\
   6178.0     &     Fe II  2374.461    & 0.52 $\pm$ 0.08   &  1.6020      \\
   6187.0     &     Fe II  2600.173    & 0.67 $\pm$ 0.10   &  1.3794      \\
   6187.5     &     Fe II  2600.173    & 0.44 $\pm$ 0.08   &  1.3798      \\
   6188.0     &     Fe II  2599.147    & 0.50 $\pm$ 0.09   &  1.3808      \\
   6201.0     &     Al III 1862.790    & 0.73 $\pm$ 0.12   &  2.3289      \\
   6202.0     &     Fe II  2382.765    & 0.94 $\pm$ 0.16   &  1.6028      \\
   6654.0     &     Mg II  2796.352    & 1.95 $\pm$ 0.11*  &  1.3796      \\
   6655.5     &     Mg II  2796.352    & 1.95 $\pm$ 0.11*  &  1.3802      \\
   6657.0     &     Mg II  2796.352    & 0.71 $\pm$ 0.08   &  1.3808      \\
   6671.0     &     Mg II  2803.531    & 1.89 $\pm$ 0.41*  &  1.3796      \\
   6673.0     &     Mg II  2803.531    & 1.89 $\pm$ 0.41*  &  1.3802      \\
   6729.0     &     Fe II  2586.650    & 0.25 $\pm$ 0.07   &  1.6016      \\
   6730.2     &     Fe II  2586.650    & 0.60 $\pm$ 0.07   &  1.6020      \\
   6732.5     &     Fe II  2586.650    & 0.18 $\pm$ 0.06   &  1.6029      \\
   6764.0     &     Fe II  2600.173    & 0.56 $\pm$ 0.10   &  1.6015      \\
   6765.5     &     Fe II  2600.173    & 0.71 $\pm$ 0.09*  &  1.6020      \\
              &     Fe II  2599.147    & 0.71 $\pm$ 0.09*  &  1.6030      \\
   6768.0     &     Fe II  2600.173    & 0.46 $\pm$ 0.09   &  1.6029      \\
   6788.0     &     Fe II  2607.866    & 0.14 $\pm$ 0.05   &  1.6030      \\
   7274.2     &     Mg II  2796.352    & 1.11 $\pm$ 0.10   &  1.6015      \\
   7275.5     &     Mg II  2796.352    & 0.83 $\pm$ 0.08   &  1.6020      \\
   7277.0     &     Mg II  2796.352    & 0.44 $\pm$ 0.07   &  1.6025      \\
   7278.5     &     Mg II  2796.352    & 0.83 $\pm$ 0.08   &  1.6029      \\
   7293.0     &     Mg II  2803.531    & 0.89 $\pm$ 0.09   &  1.6015      \\
   7294.5     &     Mg II  2803.531    & 0.98 $\pm$ 0.09   &  1.6020      \\
   7296.0     &     Mg II  2803.531    & 0.24 $\pm$ 0.06   &  1.6025      \\
   7297.0     &     Mg II  2803.531    & 1.00 $\pm$ 0.10   &  1.6029      \\
   7803.7     &     Fe II  2344.214    & 0.71 $\pm$ 0.28   &  2.3289      \\
   7931.8     &     Fe II  2382.765    & 0.46 $\pm$ 0.21   &  2.3288      \\
  16520\footnote{ Emission line.[O III] flux = (4 $\pm$ 3) $\times$ 10$^{-17}$
erg cm$^{-2}$ s$^{-1}$.}  &     [O III] 4960.295   & -(16 $\pm$  7)    &  2.3304      \\
  16679\footnote{Emission line.[O III] flux = (12 $\pm$ 3) $\times$ 10$^{-17}$
 erg cm$^{-2}$ s$^{-1}$.} &     [O III] 5008.240   & -(21 $\pm$  6)    &  2.3304      \\
  21863\footnote{Emission line.  H$\alpha$ flux = (11 $\pm$ 2) $\times$ 10$^{-17}$ erg 
cm$^{-2}$ s$^{-1}$.} &     H I    6564.610    & -(37 $\pm$ 10)    &  2.3304      \\
\noalign{
\smallskip} \hline \hline\end{tabular}
} \normalsize \rm
\label{LinesTable}
\end{center}
\end{minipage}
\end{table}

\section{Results}  
  \label{resultados}

\subsection{The blueshifted absorption line systems}  
  \label{resultados1}  

The VLT/UVES data set (Fig. \ref{SpectrumFull}) shows a
large number of absorption lines, as well as  Ly$\alpha$ in emission
(see Table \ref{LinesTable} for a list, and also \citealt{mol02b, sta05, jak05} on the Ly$\alpha$ emission line).
Some of the absorption lines are due to two foreground systems
at redshifts $z=1.3820$ and $z=1.6020$ \citep[][see \citealt{tej07, tej09, ver09} for further discussions on the foreground absorbers]{mol02b, mir03}, but the remaining lines are from
multiple systems found in the redshift range $2.2967 \leq z \leq 2.3291$.

\begin{figure}[b!]  
\begin{center} 
  \resizebox{8.0cm}{!}{\includegraphics[clip]{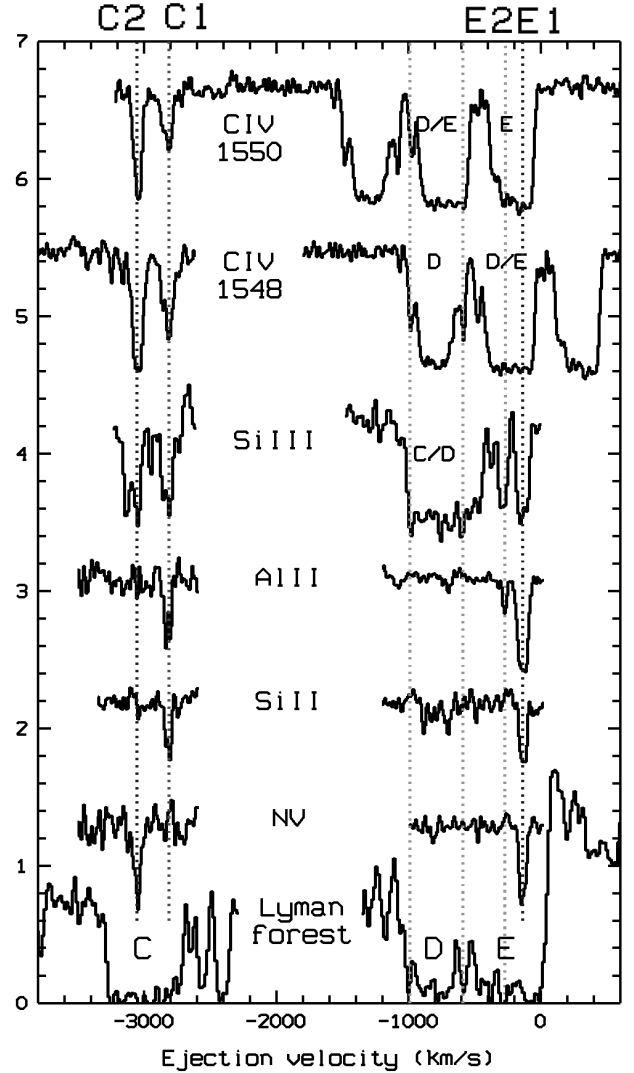}}  
 
      \caption{In this figure we have converted the observed wavelengths
               into velocities relative to the host redshift ($z=2.3304$) 
               for each ionic species significantly detected. Ly$\alpha$ 
               is plotted at the bottom. To optimize the signal-to-noise 
               ratio two lines of N {\small V} (1,238 and 1,242 \AA{}) and 
               four lines of Si {\small II} (1,260, 1,264, 1,533 and 
               1,816 \AA{}) have been coadded. The rest of the lines are 
               all single. The absorption components around $z \sim$ 2.3 
               observed at velocities up to 3,250 km s$^{-1}$ indicate the 
               presence of fast winds, ejected by a hot and massive stellar 
               progenitor. The D and E complexes both contain velocity 
               broadened, BAL-like components. Within the E complex there 
               are narrow lines, E1 and E2, presumably from the ISM of the 
               host galaxy. There are also narrow shells in the high-velocity 
               system C, C1 and C2. The dotted lines mark the positions of 
               E1, E2, the ``dips'' at the leading and the trailing edges of
               D, and the C1 and C2 systems.}  
    \label{uves2 spectrum} 
\label{Systems}
\end{center}  
\end{figure}

In Fig. \ref{Systems} we show the three wide absorption line complexes
in the redshift range $2.2967 \leq z \leq 2.3291$.
The absorption profiles are plotted
in velocity space and marked C, D, and E
following the naming convention used by  \citet{mol02a}.
The higher resolution of the UVES spectrum allowed us to identify
four narrow line systems embedded inside the velocity structure of the
broad systems C and E. We name those systems C1, C2, E1 and E2.
In the following we discuss all of those systems in more detail in order 
of increasing velocity
relative to the host redshift ($z=2.3304$, see \kref{resultados2}).

{\it E complex:} The Ly$\alpha$ absorption width translates to velocities
in the range 40--470 km s$^{-1}$, while in C IV (where we observe the
narrowest trough), the range is 40--390 km s$^{-1}$.  The difference
in the high-velocity cut-offs
(390 vs.\ 470 km s$^{-1}$) is most easily understood
as a column density effect, i.e., the column density of the
high velocity end is too low for C IV to be detectable. Alternatively
it could be caused by lower metallicity in the higher
velocity gas of system E, or a different ionization.
The simplest interpretation of the E complex is that it is caused by gas,
spanning velocities from 40--470 km s$^{-1}$ and densities decreasing from
the slowest to the fastest moving part.
Similar profiles have been observed in QSO broad absorption line (BAL)
systems, where velocities up to $\approx 0.1c$ have been found. BAL
systems are interpreted as velocity broadening due to ejection and/or
radiative acceleration. Such BAL-like absorption troughs are not
meaningfully fitted by standard (Voigt) profiles that assume a
Gaussian velocity field.

Embedded in the E complex are two narrow line
systems at 134 km s$^{-1}$ (E1) and 262 km s$^{-1}$ (E2). The strongest
component, E1,
splits equally into two subcomponents at 114 km s$^{-1}$ and
150 km s$^{-1}$ (see Fig~3)
and exhibits absorption from a large number of both high and low ionic species
(N {\small V}, C {\small IV}, Fe {\small II}, Al {\small II}, Al {\small III},
Si {\small II}, Si {\small III}, Si {\small IV}).
The metal line absorption of E1 has similarities to so-called damped
Ly$\alpha$ (DLA) systems, which in a few cases have been
identified with high redshift, star forming galaxies \citep{mol02b} 
but it has too low HI column density in this sightline to be considered a DLA.
Still, E1 is the system most likely to be identified with the
interstellar medium (ISM) in the host
galaxy of the GRB, whereby the velocity offset between the
host and the ISM ($\sim$ 134 km s$^{-1}$) represents the velocity
of the local absorbing cloud within the host. High and low ionic
species are also seen in the second system (E2), but at lower significance.
The two rightmost dotted vertical lines in Fig~2 mark the systems E1 and E2. 

{\it D Complex:}
The D complex consists of a single, wide BAL-like component seen in Ly$\alpha$,
C {\small IV}, and
Si {\small III}. Both the C {\small IV} (1,548 \AA{}) and the Ly$\alpha$
troughs give a consistent relative velocity range from
585 to 1,005 km s$^{-1}$. We find
residual flux at the bottom of both the Ly$\alpha$ and C {\small IV}
(1,548 \AA{})
troughs indicating that the gas is optically thin. We also see
significant variations in the optical thickness as a
function of relative velocity; in particular one will notice
the two sharp ``dips'' marking shells of higher column density at
the leading and the trailing edge. Those dips are seen in  Ly$\alpha$, 
in both C {\small IV} lines and in Si {\small III} (marked by dotted lines 
in Fig~2). The Si {\small III} trough is optically thick between those 
two edges, but then appears to have an optically thin extension towards 
lower ejection velocities. However, this trough overlaps with Ly$\alpha$ 
of system C, and the extension is found to belong to system C (see below). .

It is remarkable that the
D and E systems are separated in velocity by exactly the amount
required to shift the C {\small IV} 1,550 {\AA} line complex
right on top of the C {\small IV} 1,548 {\AA} one in the spectrum
\citep{mol02b,sav02}.
Such a shift is known as absorption-absorption ``line-locking'' and is
not uncommon in QSO spectra where radiative acceleration is important (e.g.,
\citealt{sri02}).
The line-locking between the D and E systems strongly suggests that a process
similar to that seen in intrinsic QSO absorption systems is indeed involved
here, i.e. radiative acceleration.

{\it C Complex:}
As discussed above, the wide Ly$\alpha$ trough of system C
is partly due to Si {\small III} absorption from system D, but also contains
Ly$\alpha$ absorption of a wide component spanning the velocity range
2,730--3,250 km s$^{-1}$ as we detect C {\small II}, C {\small IV}, 
Si {\small II}, Si {\small III}, Al {\small II} and N {\small V} 
absorption in this velocity range (Fig. \ref{Systems}). 
The finely tuned velocity offset which causes a precise overlap between
the Ly$\alpha$ (C complex) and Si {\small III} (D complex) lines could
be due to chance coincidence
(although with small probability, see below). Given the evidence for
line-locking between the two C {\small IV} lines between systems D and E,
the overlap is most
likely the result of radiative acceleration leading to line-locking
of two expanding shells corresponding to the C and D complexes.

Embedded in C we identify two narrow systems at 3,050  km s$^{-1}$ (C2)
and at $\sim$ 2,810  km s$^{-1}$ (C1). C2 is a high ionisation system
displaying strong N {\small V} and C {\small IV} absorption, but no
absorption from singly
ionized ions. C1 is a low ionization system with moderate C {\small IV}
absorption
and clear detections of the singly ionized species Al {\small II} and
Si {\small II},
which usually identifies a cloud optically thick at the Lyman limit
(see Table \ref{LinesTable}).
Most likely C2, although farthest away from the host redshift in velocity
space,corresponds to the part of the wind that is nearest to the progenitor
site in physical space. This ionization--velocity
correlation is also observed in QSOs: the high ionization
lines are often seen at the highest relative velocities for the
$z_{\rm abs}\approx z_{\rm em}$ absorption systems in QSO spectra
(e.g., \citealt{mol94}).
As for E1, C1 splits into two subcomponents, detected at 2,803 km s$^{-1}$ and
2,837 km s$^{-1}$ (low ionization) and at 2,810 km s$^{-1}$ and 2,860 km
s$^{-1}$ (high ionization). Again, we here have the ionization-velocity
correlation seen already between C1 and C2 (Fig. \ref{Systems}).
Similarly to \cite{fio05} we do consider C1 as part of the ejected systems 
simply on the basis of the low probability for it to be extrinsic, although 
we cannot exclude its association to the neighbour galaxy detected by 
{\it HST} \citep{che07}. In that case, it should be
realized that the comparatively low H {\small I} column density reported by
\citet{fyn05} would be the sum of the H {\small I} column of the two 
galaxies, such that they both have even lower H {\small I} columns.

\begin{figure}[b!]  
\begin{center} 
  \resizebox{6.0cm}{!}{\includegraphics[clip]{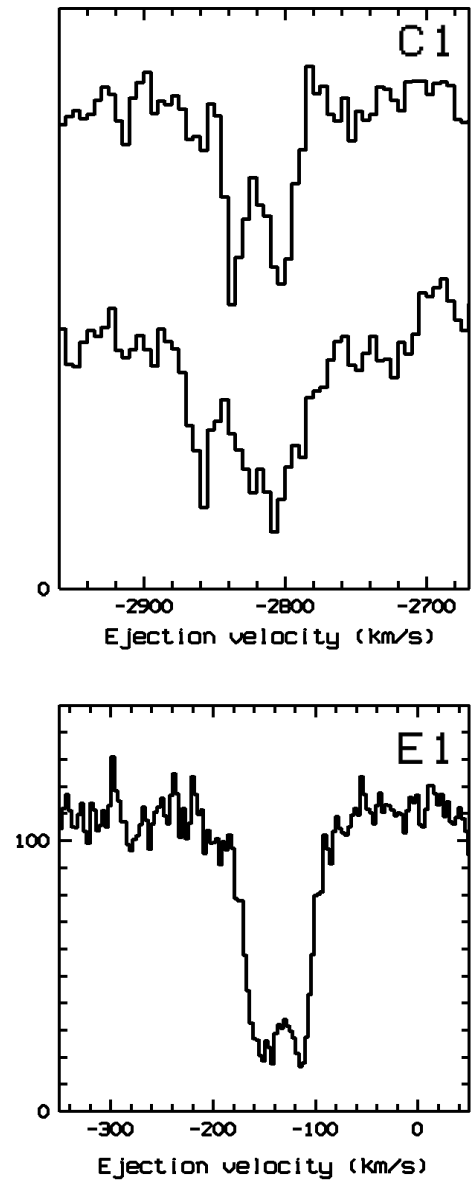}}  
 
      \caption{A zoom in of the high-resolution  VLT/UVES spectrum at
               the location of the C1 and E1 narrow-line absorption systems. 
               The data presented here are the co-add of the different ions 
               in velocity space. Thus, C1  has two subcomponents at 
               2,803 km s$^{-1}$  and 2,837 km s$^{-1}$ (low ionization, 
               Si {\small II} + Al {\small II}, top) and 2,810 km s$^{-1}$
               and 2,860 km s$^{-1}$ (high ionization, Si {\small III} + 
               C {\small IV}, bottom). The ionization--velocity correlation 
               also seen in QSOs \citep{mol94} and
               WR nebulae \citep{smi84} is noticeable. E1 equally splits 
               into two subsystems at
               114 km s$^{-1}$  and 150 km s$^{-1}$. The data shown are 
               the co-add of different ions (N {\small V} + Si {\small II} 
               + Al {\small II} + Al {\small III}) in velocity space.}  
    \label{uves3 spectrum} 
\end{center}  
\end{figure}

  \subsection{The near-IR emission lines}
	\label{resultados2}
	
\begin{figure}[t!]
\begin{center} 
  \resizebox{8.0cm}{!}{\includegraphics[clip]{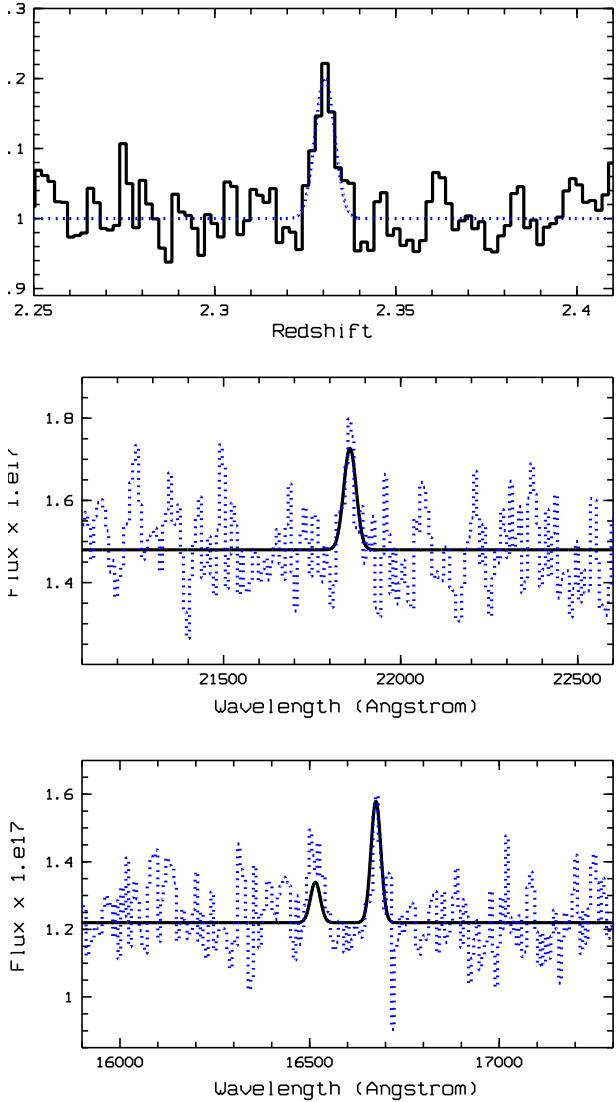}}  
 
      \caption{Bottom panels: A blow-up of the GRB 021004 nIR afterglow
      spectrum. These   VLT/ISAAC  data were  obtained  $\sim$0.6 days
      after  the GRB  and  show the H$\alpha$  and [O {\small  III}]
      emission lines arising from the host  galaxy. Top panel: the 
      co-addition of the three redshift spectra (as explained in the text) 
      is shown, providing an independent measurement of  the
      systemic redshift $z$  = 2.3304 based  on the nIR line emission.
      In   the middle and  bottom  panels  we   show the optimal  fits
      (black), and   the data  itself   (blue dotted).}   \label{uves4
      spectrum}
\label{NIRFIG}
\end{center}  
\end{figure} 

Assuming a redshift of the host of $z=2.3351$ (\citealt{mol02b}, based on
Ly$\alpha$ emission) we searched the ISAAC spectra for redshifted lines
of H {\small I} (H$\alpha$, H$\beta$) and [O {\small III}]
(4,959 \AA/5,007 \AA). There are tentative detections at low
significance of H$\alpha$ and the [O {\small III}] lines.
In order to confirm the
reality of the lines we used the method we introduced in \citet{jen01}.
Briefly the method involves rebinning the spectra into redshifts
for the rest wavelength of each expected line, in our case the lines
of H$\alpha$ and [O {\small III}] (4,959 \AA/5,007 \AA). The three
``redshift spectra'' were then coadded (with weights 2, 1 and 1 because
of the slightly higher S/N of the tentative H$\alpha$ detection). The
resulting combined line is clearly detected, as shown in Fig. \ref{NIRFIG}.

To this combined line  we fit a Gaussian  (overplotted as a blue dotted line)
and we measure the systemic redshift to be  $z=2.3304\pm0.0005$, which
is significantly different from the values derived from the Ly$\alpha$
emission  line (which  may    be  absorbed or modified    by  resonant
scattering) or from the absorption systems (which  are affected by the
local velocity of the absorbing  cloud in the  host galaxy). The width
of the Gaussian is found to be exactly the spectral resolution, so the
lines are unresolved and we can place an upper  limit on the intrinsic
line width of 600 km s$^{-1}$ FWHM.

Using a redshift of 2.3304 and the width of the resolution profile,
we now fitted Gaussian profiles to the data (with continuum level and
amplitude of the Gaussian as the only free parameters) using min-square
deviation. For the two [O {\small III} lines we further imposed a 1:3
flux ratio (\cite{ost89}), we found the following fluxes:
 flux = (4 $\pm$ 1) $\times$ 10$^{-17}$ erg cm$^{-2}$ s$^{-1}$
([O {\small III}] 4,959~\AA), (12 $\pm$ 3) $\times$ 10$^{-17}$
erg cm$^{-2}$ s$^{-1}$
([O {\small III}] 5,007~\AA) and (11 $\pm$ 2) $\times$ 10$^{-17}$ erg
cm$^{-2}$ s$^{-1}$ (H$\alpha$). The H$\beta$ flux is consistent with zero.

\section{Discussion}
\label{Discussion}

Following the suggestion that there is often a link between long-duration
GRBs and core-collapse supernovae \citep{sta03,hjo03} as proposed in
 1993 \citep{woo93}
we know that GRBs arise as the end stages of massive-star evolution.
As we describe below, the results presented here provide
independent evidence for a massive star ($\geq$ 40 M$_\odot$) progenitor
and give additional insight into the immediate surroundings and prior
evolution of the star leading up to the catastrophic event.

\subsection{A LBV / Wolf-Rayet progenitor}

The probability that the three absorption systems C1, C2, and D lie in
front of the GRB host galaxy as a chance coincidence is very small and
can be computed using the number density of intervening C {\small IV} 
absorbers  at $1.8<z<3.5$ of $1.5\pm0.2$ per unit redshift \citep{sar98}.
We find that the probability to find three of them within the redshift 
range $\Delta z = 0.0337$ spanned by C1, C2, and D is $\sim10^{-4}$.
\citet{fio05} used a similar argument but
added the column densities to obtain an even smaller probablility.
Like  \citet{fio05} we conclude that the absorbers must be intrinsic
to the host galaxy. The large blue shift then makes it unlikely that
the physical location can be anywhere else than close to the
GRB progenitor. 

Furthermore, the probability that the C {\small IV} 1,550 {\AA} line for
complex E falls on top
of the C {\small IV} 1,548 {\AA} line for complex D and the Ly$\alpha$ line
for system C falls on top of the Si III line for system D by chance is small,
of order a
few percent for each, although difficult to quantify due to the complex
profiles.
Therefore, it makes sense to seek a physical explanation for the complex
structure of the C, D, and E systems.

The multiple absorption line systems found in the spectrum of
GRB~021004 are naturally explained by multiple shell structures
formed by the stellar winds of a massive progenitor star \citep{sch03}.
Such line-driven winds are expected for very massive stars
\citep{cas75,ggs96} that end their lives as
Wolf--Rayet stars (WR) \citep{kot97}, after passing through an unstable
Luminous Blue Variable (LBV) phase \citep{lan99}. Stellar evolution
calculations
for massive stars predict a number of different phases in which the wind
properties associated with those phases, i.e., the velocities and
mass-loss rates, become strongly time-varying \citep{lan99,mae83}. The minimum
initial mass for a star to become a LBV is $\sim$ 40 M$_{\odot}$
\citep{lan94,mae00}.
During the course of their evolution, starting as O-class stars
(O $\rightarrow$ LBV $\rightarrow$ WR  $\rightarrow$ SN),
stars have distinct winds
in which their velocities are proportional to the escape velocities.
Thus, for each transition where a fast wind (i.e., when the stellar radius
was small)
follows a slow one (larger stellar radius), a shell of swept-up material
is formed. Within this simple scheme, computations of the circumstellar
medium around a 60 M$_{\odot}$ star at the zero-age main sequence
\citep{ggs96} may produce the number of distinct compressed shells in the
kinematic range seen in Fig. \ref{Systems}.

\begin{figure*}[t!]
\begin{center} 
  \resizebox{15.8cm}{!}{\includegraphics[clip]{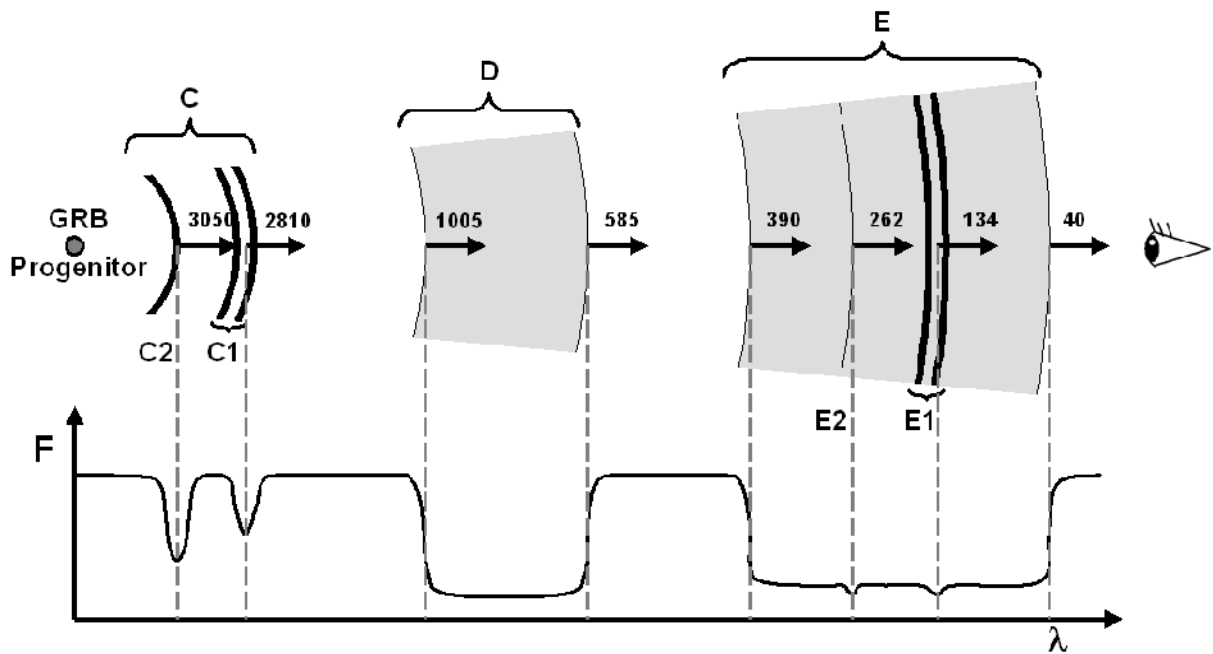}}  
 
      \caption{A sketch showing the absorption line complexes C, D and E 
               in the line of sight to GRB 021004, not drawn to scale. 
               Velocities are given in km s$^{-1}$. Complex C, with the 
               highest relative velocity, is most likely nearest to the 
               progenitor site in space and probably formed at the WR phase.
               The high-ionization sub-component C2 is the closest one to 
               the progenitor whereas the colder sub-component C1 is further 
               out and we cannot exclude that it could be even related to the
               neighbour galaxy detected by {\it HST} \citep{che07}. 
               Complex E has the lowest relative velocity, and the broad 
               part of the E complex is most likely caused by the oldest wind 
               formed when the progenitor evolved from being an O star
               into the LBV phase or in the early LBV phase. The narrow E1 
               component has the properties expected for the ISM of the host 
               galaxy. We believe that the D complex is caused by winds 
               formed during the unstable LBV phase.}  
    \label{sketch} 
\label{ToyModel}
\end{center}  
\end{figure*} 

During the LBV phase, when hydrogen shell burning occurs,
the stellar models are affected by several instabilities that cannot
yet be modelled self-consistently  and thus the unstable LBV
phase can not be fully resolved \citep{lan94}. However, several LBV observations show that
a number of different swept-up shells can be formed. In particular, the D
complex discussed above is akin to the Homunculus nebula around $\eta$~Car,
which contains $\sim$ 1 M$_{\odot}$ and expands in the range
700 and 1,000 km\,s$^{-1}$ \citep{hum94}. As the formation of each of these
swept-up shells occurs when the LBV reaches the Eddington limit
\citep{lan99}, massive stars are able to form several compressed shells before
ending their lives.

We note that, since the LBV stellar wind is asymmetric and often leads to a
bipolar or at least asymmetric nebula \citep{gru00}, the GRB has to be beamed
in the same direction as the wind for us to see the shells in absorption.
Similar, but less spectacular, complex absorption has been detected in
GRB 030226 \citep[][but see \citealt{shi06}]{klo04} and GRB 050505
\citep{ber06}; this effect is not seen, however, in several other
cases with similar quality data \citep{che07}. It is possible, therefore, that GRBs
where no such complex velocity structure is seen have a different
orientation of the wind and GRB jet.
Highly ionized gas ($\sim T=10^5$ K) like C {\small IV}, N {\small V} or Si {\small IV}, similar to
that found in the environment of GRB~021004,
has already been reported in a number of galactic ring nebulae surrounding
Wolf-Rayet stars \citep{bor97}. To form such gas
shells, forward shocks with $\sim 10^2$ km\,s$^{-1}$, or reverse shocks
with $\sim 10^2$--$10^3 $ km\,s$^{-1}$ are required.
The most natural explanation for the C complex is that it has been formed
at the WR phase, with C1 and C2 lying in an expanding free wind, before
reaching the shocked stellar wind. In fact, the
ionization-velocity correlation seen in the systems C1 and C2 has also
been detected in WR nebulae, but at lower velocities
($\sim$ 100--150 km\,s$^{-1}$; \citealt{smi84}).

In addition to the detailed picture of the spatial and velocity structure
of the GRB progenitor star at the time of explosion (Fig. \ref{ToyModel}) our observations
have also provided insight into the physical mechanism producing the
progenitor shell structure. The fact that the C1 and C2 profiles are not
P-Cygni, but
almost symmetric profiles, rules out that they are very close to
the burst (within $\sim$ 0.2 pc) and being accelerated by the GRB ionizing
flux to the observed velocities, as has been proposed elsewhere
\citep{sch03,mir03}. The detected line-locking also favours radiative
acceleration by the progenitor star.

\subsection{A starburst host galaxy}

An extremely blue host galaxy has been revealed in late time imaging
\citep{fyn05,deu05,jak05}. In order to derive the  star-forming rate (SFR), 
we have  studied the emission  lines seen in the  combined optical-nIR
spectrum.

The emission  line parameters are measured with  a Gaussian fit to the
emission line and  a flat fit to  the continuum.  If all the H$\alpha$
emission is attributed  to star  formation in  the host galaxy  we can
compute   the  SFR  as   $SFR$  (M$_{\odot}$yr$^{-1}$)  = $7.9  \times
10^{-42}\ L_{H\alpha}$ (erg/s) \citep{ken94}.   In our assumed cosmology,
the measured  H$\alpha$  intensity transforms  into  a H$\alpha$
luminosity $L_{H\alpha} =  (4.9\pm 0.9) \times  10^{42}$ erg s$^{-1}$.
This implies a SFR (without  any corrections)  of  40  $\pm$ 7
M$_{\odot}$yr$^{-1}$, which is  much larger than the present-day  rate
in our  Galaxy.    From   the  total measured  Ly$\alpha$    intensity
\citep{ajct02}, we derive  a Ly$\alpha$ luminosity $L_{Ly\alpha}$ $\leq$  
(9 $\pm$ 1) $\times$ 10$^{42}$ erg s$^{-1}$, which  implies a 
SFR $\leq$ 10 M$_{\odot}$yr$^{-1}$ in  agreement with
\citet{fyn05} and \cite{jak05}. The Ly$\alpha$/H$\alpha$ ratio of 2 
implies a low dust content in  the GRB host galaxy, as also inferred by 
\citet{deu05} and \cite{kan06} from the afterglow SED; there is also no 
excess absorption beyond the (low) Galactic value detected 
in \emph{Chandra} X-ray observations \citep{sak02a}.  
This is a large value for the SFR (see also \citet{ajct07}), although the
signal-to-noise  ratio of      the  nIR  spectrum implies  a     large
uncertainty. The  derived SFR implies  that  most of the  ongoing star
formation  is unobscured,  which is also  in agreement  with the SCUBA
results at  850 $\mu$m \citep{tan04}.

The  Ly$\alpha$   photons  produced in   the  ionized  nebula suffer a
continuous resonant scattering  in the presence of HI.  In a dust free
nebula  this mechanism   effectively  traps  the   Ly$\alpha$  photons
producing  a typical P Cygni line  profile in emission and absorption,
depending on the details of the density and velocity field.  However, if
even a little dust is  present, most Ly$\alpha$ photons will disappear
by  heating dust  grains, because  radiation absorption by  dust has a
maximum  cross section near  the  wavelength of Ly$\alpha$.  Thus, the
observed complex Ly$\alpha$ profile is  the result of both mechanisms,
resonance scattering which sets the  profile shape, and dust absorption
which further   decreases  the intensity   of  the emission  line. The
Ly$\alpha$   P Cygni  profile can  be  understood  in  terms  of these
mechanisms acting during the evolution  of the swept up HI  supershell
produced by  a massive starburst in the  host galaxy of GRB 021004. It
corresponds to stage 4 in the evolutionary scheme developed by
\citet{ten99} and \citet{mas03}; an stage compatible with GRB 021004 
being the evolution endpoint of a WR star.

\section{Conclusions}
\label{Conclusions}

We find several absorption line groups spanning a range of about 3,000
km s$^{-1}$  in velocity relative to the  redshift of the host galaxy.
The absorption profiles are  very complex with both velocity-broadened
components extending over  several 100  km  s$^{-1}$ and  narrow lines
with velocity widths  of only $\sim$ 20 km  s$^{-1}$.  By analogy with
QSO absorption line studies, the    relative velocities, widths,   and
degrees      of    ionization  of     the   lines   (``line-locking'',
``ionization--velocity correlation'')  show  that the  progenitor  had
both an extremely  strong  radiation field  and  several distinct mass
loss phases (winds). These results are consistent with GRB progenitors
being massive  stars, such as  LBVs or  Wolf--Rayet stars, and provide
further   insight into the   nature of  these   progenitors  and their
immediate environments.

The host galaxy is a prolifically star-forming galaxy at a systemic 
redshift  $z=2.3304$,  with a SFR of $\sim$40  M$_{\odot}$yr$^{-1}$ as 
also found by \citet{fyn05} and \citet{deu05}, reinforcing the potential
association of some GRB with starburst galaxies (\citet{Chri04,gor05} and
references there in).

The   {\it Swift}  mission with a predicted lifetime of ten  years
\citep{geh04} will certainly  bring  us the opportunity to   carry out
high-resolution spectroscopy for dozens of future GRBs, and to
set physical/chemical properties common to all GRB outflows.

\begin{acknowledgements}  
  
  We are grateful to M. Cervi\~no, M. M\'as-Hesse,
  R. Gonz\'alez, G. Tenorio-Tagle and S. Vergani for fruitful discussions 
  as well as the anonymous referee for the useful suggestions. 
  This research has also been partially supported by the Spanish Ministry 
  programmes AYA2004-01515, AYA 2007-06377, AYA 2009-14000-C03-01 and  
  ESP2002-04124-C03-01 (including FEDER funds). The Dark Cosmology Centre 
  is funded by the DNRFSome of the authors
  acknowledge benefits from collaboration within the EU FP5 Research
  Training Network ``Gamma-Ray Bursts: An Enigma and a Tool''.
     
\end{acknowledgements}

\end{document}